\documentclass[10pt, conference, compsocconf]{IEEEtran}
\ifCLASSINFOpdf
\else
\fi
\usepackage{amssymb}
\usepackage{graphicx}
\usepackage{subfigure}
\usepackage{mathptmx}
\usepackage{listings}
\usepackage{fancyvrb}
\usepackage{fancyhdr}
\usepackage{multirow}

\usepackage[ruled]{algorithm2e}
\usepackage{amsmath}

\begin{document}
%
\title{AxPUE: Application Level Metrics for Power Usage Effectiveness in Data Centers}


\author{
Runlin Zhou$^{1}\ \ \ $Yingjie Shi$^{2}\ \ \ $Chunge Zhu$^{1}\ \ \ $Fan Liu$^{2}$\\
\\
\begin{tabular}{c}
$^1$National Computer Network Emergency Response Technical Team\\
Coordination Center of China, Beijing, China \\
$^2$State Key Laboratory Computer Architecture, Institute of Computing\\
Technology, Chinese Academy of Sciences, Beijing, China\\
\end{tabular}
\\
\\
\begin{tabular}{cccc}
zhourunlin@cert.org.cn &shiyingjie@ict.ac.cn &jadove@163.com &liufan@ict.ac.cn\\
\end{tabular}
\\
}

%


\maketitle

\begin{abstract}
The rapid growth of data volume brings big challenges to the data
center computing, and energy efficiency is one of the most concerned
problems. Researchers from various fields are now proposing
solutions to green the data center operations. Power usage
effectiveness metric plays an important role in the energy saving
research. However, the exising usage effectiveness metrics focus on
measuring the relationship between the total facility energy
consumed and the IT equipment energy consumed, without reflecting
the energy efficiency of applications. In this paper, we analyze the
requirements of application-level metrics for power usage efficiency
of the data centers, and propose two novel energy efficiency metrics
to provide strong guidance and useful insight to data center design
and optimization. We conduct comprehensive experiments in the
practical data centers using BigDataBench, a big data benchmark
suite, and the results demonstrate the rationality and efficiency of
AxPUE in measuring the actual computation energy consumption in data
centers.

\end{abstract}


%
\IEEEpeerreviewmaketitle

\section{Introduction}

As the development of information industry, massive data has been
produced in  various applications, including online transaction
data, web access logs, sensor data, scientific data, etc. According
to IDC white book\cite{IEEEhowto:idc}, the data size will reach 35ZB
by 2020. The era of Big Data has arrived, which brings big
challenges to data centers, and energy efficiency is one of the most
important problems. The data centers and associated facilities
consumed about 1.3\% electricity of all the electricity use in the
world, and 2\% of all electricity use for the US\cite{IEEEhowto:us}.
It makes the corporations and organizations under great pressure to
reduce the operation cost of data centers, and also gets wide
attentions from both industry and academia. Researchers are
proposing various solutions to green the data center operations from
different fields.

\begin{figure}
\centering
\includegraphics[width=3.3in,height=1.6in]{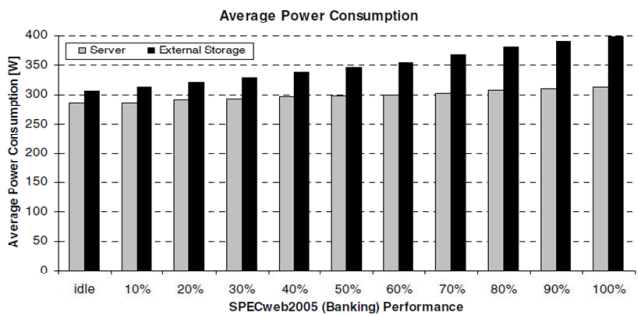}
\caption{Average Power Consumption \cite{IEEEhowto:tu}}
\label{figure:power}
\end{figure}

As the key metric for measuring infrastructure energy efficiency for
data centers, PUE(Power Usage Effectiveness) is now widely adopted
and used in the global industry. PUE was first proposed by the Green
Grid Association in 2007\cite{IEEEhowto:pue}, and its measure
methodology has been refined according to the industry feedback over
the past years\cite{IEEEhowto:pue1}\cite{IEEEhowto:pue2}. PUE is an
excellent metric to reflect how well a data center is delivering
energy to its information technology equipment, and it provides a
way to compare the operation efficiency of different data centers.
In addition, PUE is always used to evaluate the efficiency
improvement of newly proposed energy management mechanisms. However,
it focuses on illustrating the energy consumption in the
infrastructure level, without reflecting the energy efficiency of
applications and workloads. In order to evaluate the impact of
workloads on the consumed power of servers, Klaus-Dieter conducted
an experiment based on the SPECweb2005 (Banking)
workload\cite{IEEEhowto:tu}, the results are shown in Figure
\ref{figure:power}. The server power consumption ranges from ~290W
at idle to ~300W at 100\% performance, while the storage power
consumption ranges from ~310W at idle to ~390W at 100\% performance.
According to the definition of PUE\cite{IEEEhowto:pue}, it reflects
the ratio of total data center facility energy to the information
technology devices energy. Though the workload continues to
increase, the power consumed by the information technology devices
does not change significantly. The energy consumptions of other
infrastructure devices in the data center are not affected directly
by the applications, so we can infer that the PUE metric will change
little as the elements of application level change, such as data
volume, workload type, etc. What's more, suppose some software
energy efficiency optimizations are applied to the data center,
which makes the energy consumed by the IT equipment decreased
significantly, and the other infrastructure devices' energy
consumption does not change much. At this time, the PUE metric even
becomes larger, which means energy efficiency degradation. Genarally
speaking, PUE has several limitations to measure the energy
efficiency of the whole data center. Firstly, PUE measures the
energy of a data center allocated to the information technology
equipment, however, it does not provide the description of the
energy consumed by the actual computations. Though it plays a
critical role in improving the energy efficiency of non-information
technology device infrastructures, it contributes little to analyze
and improve the efficiency of information technology equipment
itself. Secondly, PUE does not reflect the applicaiton
characteristics into the energy efficiency measurement, such as
applicaiton complexity, data volume, etc. During the procedure of
data center planning with specific applicaitons and data scale, it
provides little help. An application-level comprehensive metric is
of urgent requirement in the energy optimization of data centers.

In this paper, we associate the IT equipment efficiency and the
energy consumed in the data center based on the big data application
analysis, and propose two novel efficiency metrics to provide strong
guidance and useful insight to data center design and optimization.
Our contributions include:

\begin{enumerate}
\item We analyze the requirements of application-level metrics for power usage efficiency of
the data centers, and propose two novel metrics to illustrate
intuitive description for the power consumption for the actual
computation in data centers. The proposed metrics provide meaningful
guidance for optimizing the energy consumption structure of data
centers based on specific applications.

\item We give detailed descriptions and calculation formulas for the
proposed metrics, analyze the factors that affect the metrics, and
propose the computation methodology based on big data applications.

\item We conduct comprehensive experiments in the practical data
centers based on Xeon platform using BigDataBench--a big data
benchmark suite, and the results demonstrate that AxPUE could
provide reasonable measurement for the actual computation energy
consumption.
\end{enumerate}

The rest of this paper is organized as follows. In Section 2, we
present the related work. In Section 3, we analyze the requirements
of application-level metrics in data center designing and
optimizing. In Section 4, we give the definition of AxPUE, and
describe the computation methodology. In Section 5, we calculate the
PUE and AxPUE of an actual data center, and illustrate the
experiment results. Conclusions and future work are given in Section
6.

\section{Related Work}
In general, our work in this paper is related to two fields: power
usage efficiency metric definition and computation in data centers,
energy management optimization based on the power efficiency metrics.
In order to measure the energy efficiency of data centers, the Green
Grid proposed two metrics in 2007: Power Usage Effectiveness (PUE)
and its reciprocal, Data Center Efficiency
(DCE)\cite{IEEEhowto:pue}. Since then PUE has received wide adoption
in the industry, while DCE has had limited concerns because of the
misconception of what data center efficiency really means. After
that, the Green Grid redefined the data center efficiency as data
center infrastructure efficiency(DCiE), and the confusion was
avoided\cite{IEEEhowto:dcie}. Though the definitions of PUE and DCiE
are straightforward and easy to understand, the computations are
complicated due to various influencing factors in the data centers.
As a result, a series of work was conducted to get more accurate
PUE\cite{IEEEhowto:pue1}\cite{IEEEhowto:pue2}\cite{IEEEhowto:book}.
Based on the industrial feedbacks of using PUE, several variants of
PUE were proposed: partial PUE (pPUE) was designed for the situation
in which only a particular portion of the facility was focused in
the measurement\cite{IEEEhowto:ppue}; PUE Scalability was designed
to describe how well the total energy consumption scales with
changes in IT power load\cite{IEEEhowto:ppue}. These two metrics
both focus on the IT equipment power consumption, and then as a
evolution, Data Center Productivity (DCP) was proposed to measure
the total power of the data center allocated to the $useful$ $work$
produced in the data center\cite{IEEEhowto:dce}. Based on the
various measurements of the $useful$ $work$, a family of metrics
called DCxP were proposed in \cite{IEEEhowto:proxy}. Overall, all
the existing energy efficiency metrics focus on the power
consumption analysis on the infrastructures of data centers, without
considering the application characteristics.

Based on the proposed energy efficiency metrics, much work has been
done to improve the data center power consumption. The existing work
can be classified into four categories: energy management of data
center components, such as CPU, disk,
etc.\cite{IEEEhowto:part1}\cite{IEEEhowto:part2}; energy management
of data center
systems\cite{IEEEhowto:sys1}\cite{IEEEhowto:sys2}\cite{IEEEhowto:sys3};
thermal management and
mechanisms\cite{IEEEhowto:re1}\cite{IEEEhowto:re2}\cite{IEEEhowto:re3};
power consumption management in virtual
fields\cite{IEEEhowto:vir1}\cite{IEEEhowto:vir2}\cite{IEEEhowto:vir3}.
In addition to this, some researchers proposed energy efficiency or
cost effectiveness provisioning tools for workloads in data centers
\cite{IEEEhowto:lugang}\cite{IEEEhowto:phenix}. Generally speaking,
most of the existing work focused on the energy-saving technology
from single field, there is little work on the corresponding
technology considering the interactions of technologies from
different fields. One encumbrance is the lack of metrics that can
reflect the comprehensive energy efficiency from application level
to infrastructure level of data centers.

\section{Requirements of Application Level PUE}
As described in the former sections, the existing power usage
efficiency metrics evolve from the initial metric PUE\cite{IEEEhowto:pue}, and they
all focus on the energy efficiency of IT devices and other infrastructures
of data centers. In this section, we first give brief description of
PUE, then we formulate the necessities of application level PUEs based
on two application scenarios. At last we list the criteria of
application level PUE for big data centers.

\subsection{Overview of PUE}
PUE is proposed to help data center owners and operators better
understand and improve the energy efficiency of their existing data
centers, as well as help them make better decisions on new data
center deployments. In addition, it provides a dependable way to
measure results against comparable IT organizations\cite{IEEEhowto:pue}. It is defined
as the ratio of total data center facility energy to the IT
equipment energy, just as shown in Equation 1:

\begin{eqnarray}PUE=\frac{Total \; Facility\; Energy}{IT \;Equipment \;Energy}\end{eqnarray}

During the equation, $Total$ $Facility$ $Energy$ is defined as all
the energy of the infrastructures to keep the normal operation of
the data center, which includes several components:

\begin{itemize}

\item \textbf{Power Transmission Equipment:} uninterruptible power
supply(UPS), switch devices, generator, power distribution
unit(PDU), battery, the power transmission loss before external to
the IT equipment, etc;

\item \textbf{Cooling Devices:} chiller, computer room air conditioning units
(CRACs), direct expansion air handler (DX) units, cooling towers,
pumps, etc;

\item \textbf{IT Equipment:} computer, storage, network equipment,
etc;

\item \textbf{Other Components:} lighting of data centers, etc.
\end{itemize}

\begin{figure}
\centering
\includegraphics[width=2.7in,height=2.4in]{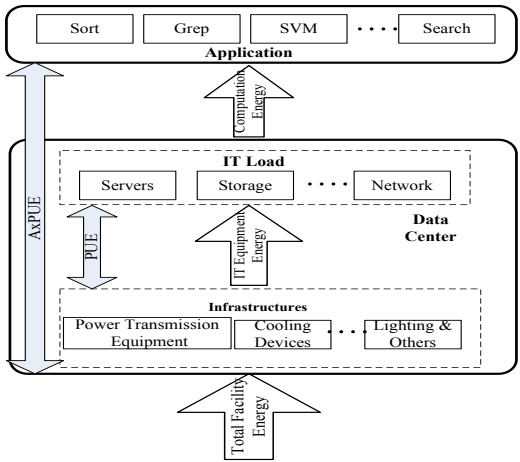}
\caption{Calculation Framework of PUE and AxPUE} \label{figure:pue}
\end{figure}

$IT$ $Equipment$ $Energy$ is defined as the energy associated with
all the IT devices and the supplemental equipment(KVM switches,
monitors, and workstations or laptops used to monitor or otherwise
control the data center). The calculation framework of PUE is
illustrated in Figure \ref{figure:pue}, where the energy structure
is divided into three hierarchies: the total facility energy
entering the data center, the energy consumed by IT equipment, and
the energy consumed by the final computation of applications. PUE
covers the lower two hierarchies, without the application elements,
which we consider plays an important role in data center energy
structure optimization and management improvement.

\subsection{Why Application Level PUE}
We begin to formulate the necessities of application level PUE by two
application scenarios.

\emph{\textbf{Scenario 1.}} The researcher of data management field
named Bob is interested in improving the performance of big data
analysis, also he concerns green data center. Recently he improves
one data classification algorithm, which has less time complexity
than the existing algorithm. He wonders whether his work makes a
contribute to the energy saving of data centers. So he runs his new
algorithm in the data classfication application of a data center,
and monitors the energy efficiency metric PUE. It seems that the PUE
has no obvious variations with the old classification algorithm, can
this result demonstrate the contribution of the new algorithm to the
data center energy saving?

\emph{\textbf{Scenario 2.}} With the expansion of business in
company C, the data to be processed and managed of the data center
increases sharply. The finance department of C has to makes annual
financial budget, and they ask the data center administrator Alice
to give a budget plan of the data center power consumption. Alice
could estimate the data volume increasement of the data center in
the next year according to the data process history and business
development trend. However, the existing data center energy
efficency metric does not consider the impact of applicaiton-level
characteristics, how to estimate the power increasement due to the
data volume increasement?

From the above two scenarios, we can find some limitations of PUE,
which measures the relationship between the total facility energy
consumed and the IT equipment energy consumed, without considering
the appication and workload. Energy saving is such a critical
problem that researchers from many fields pay great attention, such
as architecture, operating system, data mangement, etc. And the
energy efficiency problem must be solved from many aspects. However,
just as illustated in Scenario 1, PUE can not measure the
effectiveness of any changes made upon the data center
infrastructure. So a comprehensive metric that reflects the overall
energy consumption is in urgent need. Data growth is now a universal
trend, and data volume plays an imporant role in the data center
design and planning. PUE does not consider the application characteristics
in the measurement, so it provides little reference information in the data center
planning according to data scale and application complexity, just as described in Scenario 2.
How to reflect the relationship of data volume and power
consumed is also a meaningful problem.

\subsection{Criteria of AxPUE}

Based on the above analysis, we argue that appication-level power
usage effectiveness is of great importance, and its measurement design
should follow the listing criteria:

\begin{itemize}

\item \textbf{Sensitivity:} The metric should react immediately to
changes of both application characteritics and data center
facilities, and provides accurate information;

\item \textbf{Usability:} The parameters in the metric should be
easy to aquire using common collecting tools, and bring as little
influence as possible to the normal data center operations;

\item \textbf{Comparability:} The metric should adopt unified
measurement for various applications, so it can provide meaningful
insight to energy optimization of different data centers.

\end{itemize}

\section{Methodology of AxPUE}

We propose two application level power usage effectiveness metrics:
ApPUE (Application Performance Power Usage Effectiveness) and AoPUE
(Application Overall Power Usage Effectiveness), which constitute a
family of applicaiton level PUEs named AxPUE. In this section, we
first give the detailed definition of these two metrics, and then
describe the acquisiton and computation methodology.

\subsection{Definition}

\newtheorem{definition}{Definition}
\begin{definition}[ApPUE]
ApPUE is a metric that measures the power usage effectiveness of IT
equipments, specifically, how much of the power entering IT
equipmetns is used to improve the application performance.
\end{definition}

ApPUE reflects the energy efficiency of IT devices, and connected
the application characeristics and power consumed. It can be used to
evaluate the energy effectiveness provided by  IT device suppliers. ApPUE is sensitive
to application characteristics, and it's associated with a time window, which ranges from
the start of the application to the completion.
ApPUE is a ratio of application performance to the IT equipment
power during the time window, as shown in Equation 2:

\begin{eqnarray} ApPUE=\frac{Application \; Performance}{IT
\;Equipment \;Power}\end{eqnarray}

During this equation, the \emph{IT Equipment Power} is defined as
the average rate of \emph{IT Equipment Energy} consumed, and the
definition of \emph{IT Equipment Energy} is the same with that of
PUE. The \emph{Application Performance} is defiend as the data
processing performance of applications, specifically, the amount of
work done in unit time. Its detailed meaning is associated with the
application type, we will introduce the computation methodology in
the next subsection. This equation gives the computation of ApPUE
for a single application, and a typical data center may support
different applications. We give the following equation to compute
ApPUE for various applciaitons:

\begin{eqnarray} ApPUE=\sum_{i=1}^{n}{ApPUE_{i}*\omega_{i}}\end{eqnarray}

$ApPUE_{i}$ represents the application PUE value of the $i$th
application during its corresponding time window. The energy
consumption of different applications vaires, the application
consumed more energy contributes more to the energy efficiency
metric. We adopt $\omega_{i}$ to represent the weight of the $i$th
applicaiton, it is computed through the equation:

\begin{eqnarray}\omega_{i}=\frac{{IT \;Equipment \;Power}_{i}}{\sum_{i=1}^{n}{{IT
\;Equipment \;Power}_{i}}}\end{eqnarray}.

In order to measure the overall energy usage efficiency of data
centers, we propose another metric named AoPUE.

\begin{definition}[AoPUE]
AoPUE is a metric that measures the power usage effectiveness of the
overall data center system, specifically, how much of the total
facility power is used to improve the application performance.
\end{definition}

AoPUE measures the energy efficiency of the total data center
facility power to the application performance, it reflects the data
center productivity. Its computaiton formulas is shown in Equation
3:

\begin{eqnarray} AoPUE=\frac{Application \; Performance}{Total \; Facility\; Power}\end{eqnarray}

During this equation, the meaning of \emph{Application Performance}
is the same to that of ApPUE, and the meaning of \emph{Total
Facility Power} is the rate of \emph{Total
Facility Energy} used, which has the same meaning with that of PUE. AoPUE can also be
understood as the ratio of ApPUE to PUE, it can be computed through
the following equation:

\begin{eqnarray} AoPUE=\frac{ApPUE}{PUE}\end{eqnarray}

\subsection{Acquisition and Computation}

Though the definitions of ApPUE and AoPUE are straightforward and
easy to understand, their computaions are difficult due to the
complicated elements affecting the values of AxPUE. As described in
the formal subsection, there are three factors in the AxPUE
computing: \emph{IT Equipment Power}, \emph{Total Facility Power},
and \emph{Application Performance}. The acquisiton methodology of
the first two factors are widely researched in the PUE computation,
and in this paper, we focus on the acquisition of \emph{Application
Performance}.

Different from \emph{IT Equipment Power} and \emph{Total Facility
Power}, \emph{Application Performance} has direct connection to
various user applications, and itself is also a metric rather than a
pure vaule, so its computation is more complicated. The computation
methodology of \emph{Application Performance} includes two phases:
the unified performance metric definition of various applications,
and the benchmark design for big data centers.

There are various metrics to evalute applicaiton performance, such
as runtime, throughput, etc. In order to provide unified
quantitative measurement for different big data applicaitons over
various data sets, we adopt the \emph{data processing
rate}\cite{IEEEhowto:rate} as the \emph{Application Performance}
metric, and  the \emph{data processing rate} varies as different
application types. We adopt the big data center application
categories in \cite{IEEEhowto:class}, and classify the applicaitons
into three categories, and adopt different measurements according to
their application semantics, as shown in Table \ref{metric}. For
data analysis application, the metric of data processing rate is
defined as the input data size divided by the application running
time. For service and interative real-time applications, the metric
is defined as the number of requests or transactions completed in
unit time. And the metric for high performance computing is defined
as the number of float operations in unit time.

\begin{table*}
\centering \caption{Application Performance Metric}
\begin{tabular}{|c|c|c|}
\hline Application Category&Examples&Metric\\
\hline Service Application&Search engine, Ad-hoc queries& Number of requests answered in unit time\\
\hline Data Analysis Application&Data mining, Reporting, Decision support, Log analysis&Volume of data processed in unit time\\
\hline Interative Real-time Application&E-commerce, Profile data management&Number of transactions completed in unit time\\
\hline High Performance Computing&Scientific computing&Number of floating-point operations in unit time\\
 \hline\end{tabular} \label{metric}
\end{table*}\

The data centers support a diversity of data processing
applications, and different data centers focus on different
application types. In order to get the reasonable and comparable
\emph{Application Performance} metric, the benchmarks should satisfy
the following conditions: first, the workloads in the benchmark
should be representive for big data applications, and the
composition of different workloads should be adjustable to represent
different application types; secondly, a scalable data generation
tool should be provided in the benchmark, so a relatively stable
performance metric can be gotten from the
experiments\cite{IEEEhowto:data}. In this paper, we adopt the
BigDataBench\cite{IEEEhowto:bigben} to acquire the \emph{Application
Performance} of a specified big data center. BigDataBench is a big
data benchmark suite open-sourced recently and publicly
available\cite{IEEEhowto:bigbenweb}, and all the above conditions
are well fullfilled in BigDataBench.

\section{Experiments}

In this section, we compute the AxPUE and PUE of three
representative categories of big data workloads: comprehensive big
data application, single big data application, and high performance
computing application which makes the CPU and memory of every server
fully used. The experiment results demonstrate that AxPUE can well
reflect the energy usage efficiency for applications.

\subsection{Experiment Overview}

\begin{table}
\caption{Server Configurations}\label{conf} \center
\begin{tabular}{|c|c|}
  \hline
  CPU Type & Intel \textregistered Xeon E5645\\ \hline
  \# Cores & 6 cores@2.4G \\ \hline
  \# threads& 12 threads \\ \hline
    \#Sockets & 2 \\ \hline
  \hline
  L1 DCache & 32KB, 8-way associative, 64 byte/line \\ \hline
  L1 ICache & 32KB, 4-way associative, 64 byte/line \\ \hline
  L2 Cache & 256 KB, 8-way associative, 64 byte/line \\ \hline
  L3 Cache &  12 MB, 16-way associative, 64 byte/line \\ \hline
  Memory & 32 GB , DDR3 \\  \hline
    Network & 1 Gb ethernet link\\ \hline
\end{tabular}
\end{table}

\begin{table}
\centering \caption{Workloads}
\begin{tabular}{|c|c|c|}
\hline Workload&Data Size&Description\\
\hline BigDataBench&100GB&Comprehensive Workload\\
\hline SVM&20GB&Single Workload\\
\hline Sort&100GB&Single Workload\\
\hline Grep&100GB&Single Workload\\
\hline Linpack&32GB&HPC Workload\\
 \hline\end{tabular} \label{workload}
\end{table}\

\begin{table*}
\centering \caption{Experiment Results}
\begin{tabular}{|c|c|c|c|c|c|c|}
\hline Workload&IT Equipment Power&Total Facility
Power&Application Performance&PUE&ApPUE&AoPUE\\
\hline
BigDataBench&100.412KW&147.323KW&563.271KB/s&1.467&5.6096&3.823\\
\hline SVM&103.766KW&150.897KW&134.854KB/s&1.454&1.2996&0.894\\
\hline Sort&92.122KW&138.481KW&1588.128KB/s&1.503&17.2394&11.468\\
\hline Grep&92.331KW&138.636KW&24916.998KB/s&1.502&269.866&179.730\\
\hline Linpack&122.679KW&170.685KW&50.46GFLOPS&1.391&0.411&0.295
\\
 \hline\end{tabular} \label{res}
\end{table*}\

In our data center, there are 18 racks which contain 362 servers, 3
large switches, and 8 switches with 28 ports. We sample 8 servers
from the whole cluster to conduct our experiments. The details of
configuration parameters of each server are listed in Table
\ref{conf}. The applications we adopt are listed in Table
\ref{workload}, and all the applications are performed on Hadoop
1.0.2. We adopt four typical data analysis workloads from the
BigDataBench workload to construct a comprehensive application,
which includes sort, wordcount, grep and bayes. We also adopt three
single workloads: SVM, Sort and Grep. The Linpack workload is used
to make a full use of the CPU and memory of servers. We conduct two
experiments to demonstrate the effectiveness of AxPUE: the first
experiment is executed on all the workloads to reflect the variation
trend of AxPUE on different applicaitons, and the second experiment
is executed on specified workload with different implementations and
data scales.

\subsection{Experiment on Different Workloads}

\begin{figure}
\centering
\includegraphics[width=2.7in,height=1.8in]{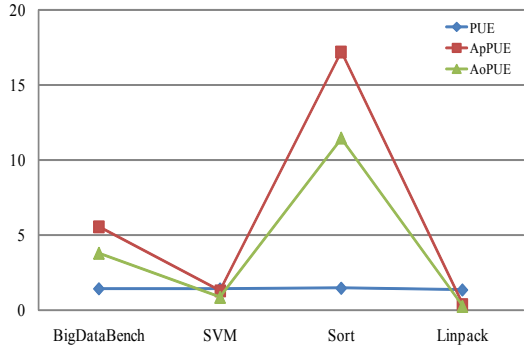}
\caption{Variation Trends of PUE and AxPUE} \label{figure:res}
\end{figure}

In the framework of Hadoop, there are one master node and serveral
slave nodes, we acquire the server power consumption from all the
nodes of our Hadoop cluster. The IT equipment power contains server
power and the swithgear power, we monitor the power consumption of
the devices for every application. The facility power consumption
except IT equipment is computed through PDU, UPS and cooling
devices. Table \ref{res} shows the detailed experiment results of
every application, Figure \ref{figure:res} illustrates the trends of
three different power usage effectiveness metrics, because the
corresponding values of grep is too big to reflect the variation
details of the trend, we show the results of other four workloads.

The computation and data processing characteristics of all the
applications differ greatly: Grep, Sort and Wordcount scan all the
data and conduct simpler computation; while Bayes and SVM classify
data based on statistical model, and their computings are more
complicated. PUE measures the relationship between the total
facility energy consumed and the IT equipment energy consumed, so it
does not reflect the workload variation directly. From the results
we can see that, the total facility energy and IT equipment energy
varies to some extend with the applicaiton type, and the PUE values
change in a narrow range from 1.391 to 1.503. PUE does not reflect
the workload variations nor the productivity of the data center.
During our experiments, the five workloads adopt different metrics
for the application performance: Linpack adopts the float-pointing
operation rate as the metric, and the other four workloads adopt the
data processing capacity as the metric. We analyze the AxPUE values
of the four workloads that use the data processing rate as the
performance metric. The maxmum of ApPUE is 269.866 KB/J (Grep), and
the mixmum of ApPUE is 1.2996 KB/J (SVM). From Table \ref{res} we
can see that both Grep and Sort consume less IT equipment energy.
The Grep workload is the simplest during all the workloads, it scans
the whole data set and finds the records matched the key word, and
then output them, its data processing capacity is the biggest.
Though the computation of Sort is not complicated either, it has to
output all the records in the data set as the final result. What's
more, during the shuffle phase of the MapReduce processing
framework, the whole data set is written into local files of mapper
nodes and transferred to reducer nodes, then read into memories of
the reducer nodes, so the I/O load is heavy, and its data processing
capacity is much less than Grep. The IT equipment power of SVM is
also higher than the simpler workloads, and its data processing
capacity is the smallest, so its ApPUE is the smallest. The workload
of BigDataBench includes both simple applications and complicated
applications, its ApPUE is about 5 times bigger than that of SVM.
The IT equipment power of Linpack is the largest of all the
workloads, it includes complicated scientific computations and makes
both the CPU and memory in full use. The application performance
metric of Linpack is different from other workloads, so we cannot
make simple horizontal comparison for them. ApPUE and AoPUE measure
the energy efficiency from two aspects: ApPUE reflects relationship
between application and IT equipment, and AoPUE reflects the energy
consumed to the actual computaion work of the whole data center.

\subsection{Experiment on Sort}

In this experiment, we compute the AxPUE of sort with two
implementations to demonstrate that AxPUE can reflect the influence
of mechanisms of application level on the power usage effectiveness,
we also show the variation trends of AxPUE on data sets of different
sizes. There are different algorithms to implement sort on Hadoop.
In the first experiment, we adopt the sort algorithm with random
sample during the partition phase, we call it sort1. In sort1, the
mapper scan the block arranged to it, and output the immediate
results after sorting them, then these immediate results are
partitioned to different reducers. The reducers are responsible for
continous but nonoverlapping partitions. In order to keep the load
balance of all the reducers, sort1 determines the partition range of
every reducer based on random samples from the input data. In Sort2,
we set the number of reducers to be one, which means that the
outputs of all the mappers are merged in one reducer, and there is
no need of a partition method.

\begin{figure}
\centering
\includegraphics[width=2.7in,height=1.8in]{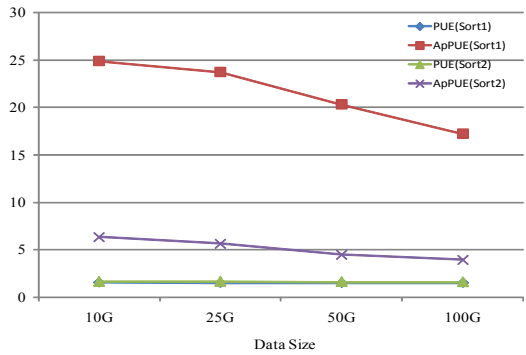}
\caption{PUE and AxPUE on Sort} \label{figure:res1}
\end{figure}

From the experiment results shown in Figure \ref{figure:res1}, we
can find that the values of PUE are almost the same of sort1 and
sort2, and ApPUE can reflect the differences of these two
implementation mechanisms. The partition method of sort1 samples
some data randomly from the input data sets, and determines the
range of every reducer based on the sampling data. The sampling
phase may cost some time and energy, however, it guarantee the load
balance of every reducer to a great extent. Though sort2 does not
need to do the sampling and the partitioning, all the immediate
results are sent to one reducer, which makes the reducer node
consumes more power, and the running time longer. So we can see from
the results that the ApPUE of sort1 is higher, and it means that
sort1 is more energy-efficient than sort2. We can conclude that
AxPUE can reflect the energy efficiency of mechanisms in the
application level, and provides meaningful guidance for the data
center energy optimization from different fields.

As the data size changes, we can see that the PUE value still
remains almost the same, it does not include the data volume factor
in the metric, so it can not provide reference information for data
center planning with specified data size. The ApPUE reflects the
data processing rate of unified energy, however, it is not
reasonable to compute the required energy of specified data volume
by extending a single ApPUE value linearly. From the results we can
observe that as the data size changes, the ApPUE does not remain the
same. As analyzed in \cite{IEEEhowto:data}, sort is an I/O intensive
application, and I/O operation is a bottleneck for Sort. When
conduct the data center planning, we suggest that the administrators
analyze the application characteristics and the variation trend of
ApPUE on different data sets, instead of scaling out a single AxPUE
with the data volume.

\section{Conclusion}

In this paper, we analyze the requirements of application-level
energy effectiveness metrics AxPUE in the data centers, and propose
two novel applicaiton-level metrics to measure the energy consumed
to the final computing application. Our experiment results
demonstrate that ApPUE and AoPUE can reflect the data center energy
efficiency variation of different workloads, and provide meaningful
guidance for greening data centers from various fields. In general,
the concept of AxPUE proposed in this work is fundamental to the
data center energy efficiency metric. ApPUE and AoPUE gives an
preliminary exploration on the application-level PUEs, we believe
that more metrics would join the family of AxPUE in the future work,
and provide meaningful guidance for energy consumption optimization
associated with appliation characteristics.

\section*{Acknowledgment}

We are very grateful to anonymous reviewers. We also thank Lei Wang
for the informative discussions that guide this work, and Jie Lv for
help in the experiments. This work is supported by the Chinese 973
project (Grant No.2011CB302502), the NSFC project (Grant
No.60933003, 61202075), the BNSF project (Grant No.4133081), and the
242 project (Grant No.2012A95).



%

\end{document}